\documentclass[10pt,conference]{IEEEtran}
\IEEEoverridecommandlockouts

\usepackage[T1]{fontenc}
\usepackage[dvipsnames]{xcolor}
\usepackage{float}
\usepackage{inputenc}
\usepackage{amsthm}
\usepackage{thmtools}
\usepackage{mathtools}
\usepackage{physics}
\usepackage{afterpage}
\usepackage{bm} % bold math
\usepackage{bbm} % black board math
\usepackage{braket}
\usepackage{dsfont}
\usepackage{amsmath}
\usepackage{amssymb}
\usepackage{tikz}
\usetikzlibrary{arrows, positioning, calc, decorations.pathreplacing}
\usepackage{tikz-cd}
\usepackage{ulem}
\usepackage[linesnumbered,ruled]{algorithm2e}
\SetArgSty{textnormal}

\usepackage{etoolbox}
\apptocmd{\sloppy}{\hbadness 10000\relax}{}{}

\usepackage{silence}
        \DeclareMathOperator{\id}{id} % identity
        \DeclareMathOperator{\diag}{diag} % diagonal matrix
        
        \newcommand*{\N}{\mathbb{N}}

        \newcommand*{\R}{\mathbb{R}}
        \newcommand*{\C}{\mathbb{C}}
        \newcommand*{\hil}{\mathcal{H}}
        \newcommand\scriptin{\raisebox{0.15ex}{$\scriptscriptstyle\in$}} % "element of" symbol for usage in sub- and superscripts
        
        \newcommand*{\one}{\mathds{1}} % identity operator
\usepackage[colorlinks]{hyperref}
\hypersetup{
    colorlinks  = true,
    citecolor   = PineGreen,
    linkcolor   = MidnightBlue,
    urlcolor    = PineGreen
}

\declaretheorem[style=plain]{theorem}

\declaretheorem[style=plain,sibling=theorem]{lemma}

\begin{document}

\title{Practical lower bounds for hybrid quantum interior point methods in linear programming
\thanks{This work was supported by the Federal Ministry for Research, Technology and Space (BMFTR) under grant number 13N17249.}}

\author{\IEEEauthorblockN{Lennart~Binkowski\IEEEauthorrefmark{1}}
\IEEEauthorblockA{\IEEEauthorrefmark{1}Strangeworks, Austin, TX 78746, United States\\
Email: lennart@strangeworks.com}}

\maketitle

\begin{abstract}
    Quantum interior point methods (QIPMs) promise polynomial speed-ups over classical solvers for linear programming by outsourcing the solution of Newton linear systems to quantum linear solvers (QLSAs).
    However, asymptotic speed-ups do not necessarily translate to practical advantages on realistic problem instances.
    In this work, I evaluate whether practical advantage of a standard hybrid QIPM pipeline can already be excluded relative to the classical open-source solver HiGHS on a broad and diverse collection of LP instances spanning eight problem families, including public benchmark libraries, such as MIPlib, and relaxations of combinatorial optimisation problems.
    Following the hybrid benchmarking paradigm initiated by Cade \textit{et al.}, I derive rigorous lower bounds on the quantum runtime under a series of highly benevolent assumptions and compare them against classical runtimes.
    I equip the QIPMs with the best-performing functional QLSA, the Chebyshev-based method, as identified by Lefterovici \textit{et al.}, and evaluate two Newton system formulations proposed by Mohammadisiahroudi \textit{et al.}: the modified normal equation system and the orthogonal subspace system.
    The exclusion analysis yields a consistent negative picture: across all instances and for any realistic quantum cycle duration, the quantum runtime lower bounds already exceed the classical runtimes, establishing that these hybrid QIPMs will offer no practical advantage over good classical solvers for realistic linear programming instances.
\end{abstract}


\section{\label{section:Introduction}Introduction}

Quantum interior point methods (QIPMs) promise polynomial speed-ups over their classical counterparts by outsourcing the solution of the Newton linear systems to quantum linear solvers.
However, asymptotic speed-ups do not necessarily translate to practical advantages on realistic problem instances.
In this work, I consider QIPMs for linear programming and compare them against state-of-the-art classical solvers on a broad and diverse set of LP instances, establishing that even under highly optimistic assumptions, the quantum methods offer no practical advantage.
This work does not attempt to simulate realistic end-to-end performance of quantum interior point methods.
Instead, it establishes instance-wise lower bounds on the runtime of a standard hybrid QIPM pipeline under a sequence of assumptions deliberately favourable to the quantum method.
The purpose is therefore one-sided: to test whether practical quantum advantage can already be excluded even before accounting for realistic iteration counts or error correction overhead.

Two prior studies have investigated the practical viability of quantum optimisation algorithms for linear programming via concrete runtime analysis.
Ammann \textit{et al.}~\cite{Ammann2023RealisticRuntimeAnalysisForQuantumSimplexComputation} benchmarked Nannicini's quantum simplex method~\cite{Nannicini2024FastQuantumSubroutinesForTheSimplexMethod} against classical simplex solvers and ruled out practical quantum speed-ups on a well-curated set of LP instances.
Dalzell \textit{et al.}~\cite{Dalzell2023EndToEndResourceAnalysisForQuantumInteriorPointMethodsAndPortfolioOptimization} performed a detailed end-to-end resource analysis of QIPMs for (non-linear) portfolio optimisation problems.
This work extends and complements both:
unlike~\cite{Ammann2023RealisticRuntimeAnalysisForQuantumSimplexComputation}, I benchmark QIPMs rather than quantum simplex methods;
unlike~\cite{Dalzell2023EndToEndResourceAnalysisForQuantumInteriorPointMethodsAndPortfolioOptimization}, I evaluate across a diverse collection of LP instances spanning multiple problem families (including Netlib, MIPlib, Max Flow, Clique, Independent Set, Vertex Cover, and StochLP).
Notably, the problems considered in~\cite{Dalzell2023EndToEndResourceAnalysisForQuantumInteriorPointMethodsAndPortfolioOptimization} are second-order cone programs, for which both the quantum and classical IPM theory is less mature than for LPs.
Linear programs therefore represent the setting where both sides are most developed and the comparison is most meaningful---yet the negative results of this work persist.
Furthermore, I equip the QIPMs with the Chebyshev-based quantum linear solver identified by Lefterovici \textit{et al.}~\cite{Lefterovici2026BeyondAsymptoticScalingComparingFunctionalQuantumLinearSolvers} as the best-performing functional quantum linear solver, and I compare two different Newton system formulations---the modified normal equation system proposed in~\cite{Mohammadisiahroudi2025ImprovementsToQuantumInteriorPointMethodForLinearOptimization} and the orthogonal subspace system proposed in~\cite{Mohammadisiahroudi2025AnInexactFeasibleInteriorPointMethodForLinearOptimizationWithHighAdaptabilityToQuantumComputers}---to assess whether the choice of classical reformulation affects the conclusions.

The benchmarking methodology follows the paradigm pioneered by Cade \textit{et al.}~\cite{Cade2023QuantifyingGroverSpeedUpsBeyondAsymptoticAnalysis}:
I derive rigorous lower bounds on the quantum runtime under benevolent assumptions and compare them against classical runtimes.
If the lower bound already exceeds the classical runtime, no refinement of the quantum algorithm can yield a practical advantage on that instance.
The results confirm a consistent negative picture: across the vast majority of instances and for realistic quantum cycle durations, even the quantum-favourable lower bound for the studied QIPM pipeline does not undercut the classical runtime.

\section{\label{section:PreviousWork}Previous work}

Roughly speaking, there are two major families of classical algorithms for linear programming (LP): derivatives of Dantzig's revolutionary simplex method~\cite{Dantzig1990OriginsOfTheSimplexMethod} and interior point methods (IPMs), pioneered by Karmarkar~\cite{Karmarkar1984ANewPolynomialTimeAlgorithmForLinearProgramming}.
Both algorithmic classes provide comparable performance in practice, despite the IPMs' exponentially better theoretical worst-case runtime.
There have been several improvements to Dantzig's initial simplex algorithm:
most notably, the revised simplex method~\cite{Wagner1957AComparisonOfTheOriginalAndRevisedSimplexMethods}, which can also handle sparse input formats, and the dual simplex method~\cite{Lemke1954TheDualMethodOfSolvingTheLinearProgrammingProblem}, which synergises well with warm-starts in branch-and-bound applications.
For example, the open source software HiGHS~\cite{Huangfu2017ParallelizingTheDualRevisedSimplexMethod}, utilised for benchmarking within this work, incorporates both enhancements in a dual revised simplex solver.
Similarly, Karmarkar's algorithm has been superseded by more performant IPMs.
The most promising implementations---such as the one within the HiGHS package~\cite{Zanetti2025AFactorisationBasedRegularisedInteriorPointMethodUsingTheAugmentedSystem}---are based on primal-dual path-following IPMs, originating from the works of Renegar~\cite{Renegar1988APolynomialTimeAlgorithmBasedOnNewtonSMethodForLinearProgramming} and boosted in practice by Mehrotra's predictor-corrector algorithm~\cite{Mehrotra1992OnTheImplementationOfAPrimalDualInteriorPointMethod}.
Both, simplex algorithms and IPMs, require subroutines for solving linear systems of equations (LSEs).
These can, e.g., be direct factorisation methods like the LU decomposition for general matrices and the faster Cholesky decomposition for Hermitian, positive-definite matrices, or iterative (Krylov) algorithms like conjugate gradient for positive-semidefinite matrices.

In essence, quantum variants of the simplex method~\cite{Nannicini2024FastQuantumSubroutinesForTheSimplexMethod} and IPMs, such as the pioneering work by Kerenidis and Prakash~\cite{Kerenidis2020AQuantumInteriorPointMethodForLpsAndSdps} and subsequent refinements for feasibility preservation~\cite{Mohammadisiahroudi2025AnInexactFeasibleInteriorPointMethodForLinearOptimizationWithHighAdaptabilityToQuantumComputers} and preconditioning~\cite{Wu2025APreconditionedInexactInfeasibleQuantumInteriorPointMethodForLinearOptimization}, outsource solving the LSEs to quantum linear solvers, while keeping the remaining steps classical, such as updating the LSEs to be solved.
The most prominent quantum linear solver is arguably the pioneering Harrow-Hassidim-Lloyd (HHL) algorithm~\cite{Harrow2009QuantumAlgorithmForLinearSystemsOfEquations}.
It is one example of the class of functional quantum linear solvers, which directly implement the (approximate) inverse of the system's matrix on the quantum computer.
This class contains further improvements to the HHL algorithm such as two approaches based on Fourier and Chebyshev representations~\cite{Childs2017QuantumAlgorithmForSystemsOfLinearEquationsWithExponentiallyImprovedDependenceOnPrecision}, and the quantum singular value transform~\cite{Gilyen2019QuantumSingularValueTransformationAndBeyondExponentialImprovementsForQuantumMatrixArithmetics}.
In contrast, adiabatic quantum linear solvers~\cite{Suba2019QuantumAlgorithmsForSystemsOfLinearEquationsInspiredByAdiabaticQuantumComputing} typically encode the solution to an LSE as ground state of a target Hamiltonian and approximate it by evolving the ground state of an initial Hamiltonian by a sufficiently slowly varied interpolation between initial and target Hamiltonian.
A recent study~\cite{Lefterovici2026BeyondAsymptoticScalingComparingFunctionalQuantumLinearSolvers}, comparing the quantitative query counts of the four mentioned functional quantum linear solvers, identified the Chebyshev-based method as the best-performing alternative.
Accordingly, I equip the quantum interior point methods with this quantum linear solver to allow for a comparison.

The benchmarking technique of combining quantitative complexity bounds for quantum methods with classical emulation has been pioneered by Cade \textit{et al.}~\cite{Cade2023QuantifyingGroverSpeedUpsBeyondAsymptoticAnalysis} and has since been utilised several times to refute manifestation of asymptotic speed-ups of certain quantum routines over classical alternatives on realistic instances.
For example, a recent study of asymptotically favourably scaling quantum SDP methods for QUBO relaxations~\cite{GSLBrandao2022FasterQuantumAndClassicalSdpApproximationsForQuadraticBinaryOptimization} excluded the possibility of any practical speed-up over existing classical methods~\cite{Henze2025SolvingQuadraticBinaryOptimizationProblemsUsingQuantumSdpMethodsNonAsymptoticRunningTimeAnalysis}.
Furthermore, Ammann \textit{et al.}~\cite{Ammann2023RealisticRuntimeAnalysisForQuantumSimplexComputation} benchmarked Nannicini's quantum simplex method against conventional simplex methods on a variety of realistic LP instances, ruling out practical quantum speed-ups entirely.
In this work, I reutilise their well-curated set of benchmark instances, since they contain a mixture of (theoretically) favourable and challenging instances for QIPMs.

\section{\label{section:Preliminaries}Preliminaries}

\subsection{\label{subsection:InteriorPointMethodsForLinearProgramming}Interior point methods for linear programming}

Consider a generic linear optimisation problem in standard form, given by
\begin{align}\label{equation:LinearProgram}
    \begin{split}
        \min_{\bm{x} \scriptin \R^{n}} &\bm{c}^{T} \bm{x} \\
        \text{s.t. } & A \bm{x} = \bm{b}, \\
        & \bm{x} \geq 0.
    \end{split}
\end{align}
Here, $\bm{c} \in \R^{n}$ is the objective vector, $A \in \R^{m \times n}$ together with $\bm{b} \in \R^{m}$ encode $m \leq n$ equality constraints, and the positivity constraint $\bm{x} \geq 0$ applies component-wisely.
Furthermore, I assume that none of the $m$ equality constraints is redundant, meaning that $A$ has full row rank.

Following standard duality theory for linear programs~\cite{Roos2005InteriorPointMethodsForLinearOptimization}, the optimal value of \eqref{equation:LinearProgram}---provided its existence---is equal to the optimal value of the dual problem
\begin{align}\label{equation:DualProgram}
    \begin{split}
        \max_{(\bm{y}, \bm{s}) \scriptin \R^{m + n}} &\bm{b}^{T} \bm{y} \\
        \text{s.t. } & A^{T} \bm{y} + \bm{s} = \bm{c}, \\
        & \bm{s} \geq 0.
    \end{split}
\end{align}

Conceptually, primal-dual path-following IPMs are initialised in a triple of primal-dual variables $(\bm{x}, \bm{y}, \bm{s})$ and iteratively update those variables via linear Newton systems for the updates $(\Delta \bm{x}, \Delta \bm{y}, \Delta \bm{s})$:
\begin{subequations}\label{equation:NewtonSystem}
    \begin{align}
        A \Delta \bm{x} &= \bm{b} - A \bm{x}, \label{equation:NewtonSystem1} \\
        A^{T} \Delta \bm{y} + \Delta \bm{s} &= \bm{c} - A^{T} \bm{y} - \bm{s}, \label{equation:NewtonSystem2} \\
        x_{i} \Delta s_{i} + s_{i} \Delta x_{i} &= \beta \mu - x_{i} s_{i}\quad \forall\, i \in [n]. \label{equation:NewtonSystem3}
    \end{align}
\end{subequations}

There are now several alternatives to solving \eqref{equation:NewtonSystem} for obtaining a Newton step.
These alternatives might, depending on the concrete LP at hand, be faster or more accurately solvable.
Here, I will focus on two particular ones: the (modified) normal equation system and the orthogonal subspaces system.
For both systems, it will be useful to introduce the notation $X \coloneqq \diag(\bm{x})$, $S \coloneqq \diag(\bm{s})$, and $\bm{1} \coloneqq (1, \ldots, 1)^{T}$.

First, I consider the normal equation system (NES) for the $m$-dimensional $\Delta \bm{y}$ update:
\begin{align}\label{equation:NormalEquationSystem}
    M \Delta \bm{y} = \bm{\sigma},
\end{align}
where
\begin{align*}
    D &\coloneqq X^{1 / 2} S^{-1 / 2} \in \R^{n \times n}, \\
    M &\coloneqq A D^{2} A^{T} = A X S^{-1} A^{T} \in \R^{m \times m}, \text{ and} \\
    \bm{\sigma} &\coloneqq A D^{2} \bm{c} - M \bm{y} - \beta \mu A S^{-1} \bm{1} + \bm{b} - A \bm{x} \in \R^{m}.
\end{align*}
Solving the NES instead of \eqref{equation:NewtonSystem} has two advantages:
First, for $m \ll n$, it is substantially smaller.
Second, the constraint matrix $M$ is, by construction, symmetric and---since $A$ is assumed to have full row rank and $(\bm{x}, \bm{s}) > 0$ holds---also positive definite.
Such systems can be solved faster than generic linear systems, e.g., by Cholesky factorisation or conjugate gradient methods \cite{Carson2024TowardsUnderstandingCgAndGmresThroughExamples}.
After obtaining the solution $\Delta \bm{y}$ to \eqref{equation:NormalEquationSystem}, the remaining updates are calculated via
\begin{align}\label{equation:NormalEquationSystemUpdates}
    \begin{split}
        \Delta \bm{s} &= \bm{c} - A^{T} \bm{y} - \bm{s} - A^{T} \Delta \bm{y} \text{ and} \\
        \Delta \bm{x} &= \beta \mu S^{-1} \bm{1} - \bm{x} - D^{2} \Delta \bm{s}.
    \end{split}
\end{align}
Therefore, while the NES is typically faster to solve, it requires additional pre- and postprocessing for obtaining the actual Newton steps.

When solving the NES inexactly, but assuming the pre- and postprocessing to be exact, a direct calculation~\cite{Mohammadisiahroudi2024EfficientUseOfQuantumLinearSystemAlgorithmsInInexactInfeasibleIpmsForLinearOptimization} shows that the resulting updates $\widetilde{\Delta \bm{y}}$ and $\widetilde{\Delta \bm{s}}$ still obey \eqref{equation:NewtonSystem2} and \eqref{equation:NewtonSystem3} while the errors accumulate in \eqref{equation:NewtonSystem1}.
A common method for improving the performance of the IPM is to transfer the residuals from \eqref{equation:NewtonSystem1} to \eqref{equation:NewtonSystem3}, requiring additional pre- and postprocessing steps~\cite{AlJeiroudi2008ConvergenceAnalysisOfTheInexactInfeasibleInteriorPointMethodForLinearOptimization}.
Since $A$ has full row rank, one can choose a subset $B$ of $m$ linearly independent columns of $A$ and consider the invertible matrix $A_{B} \coloneqq [\bm{a}_{j}]_{j \scriptin B} \in \R^{m \times m}$.
$B$ can be chosen statically or dynamically after every iteration of the IPM.
Once fixed, consider the preprocessed problem with constraint matrix $\hat{A} \coloneqq A_{B}^{-1} A$ and constraint vector $\hat{b} \coloneqq A_{B}^{-1} b$.
The modified NES (MNES) is given by
\begin{align}\label{equation:ModifiedNormalEquationSystem}
    \hat{M} \bm{z} = \hat{\bm{\sigma}},
\end{align}
where
\begin{align*}
    D_{B} &\coloneqq X_{B}^{1 / 2} S_{B}^{-1 / 2}, \\
    \hat{M} &\coloneqq D_{B}^{-1}\hspace*{-1pt} A_{B}^{-1}\hspace*{-1pt} M \big(D_{B}^{-1}\hspace*{-1pt} A_{B}^{-1}\hspace*{-1pt} \big)^{T} = D_{B}^{-1}\hspace*{-1pt} \hat{A} D^{2} \big(D_{B}^{-1}\hspace*{-1pt} \hat{A}\big)^{T}, \text{ and} \\
    \hat{\bm{\sigma}} &\coloneqq D_{B}^{-1} A_{B}^{-1} \bm{\sigma} = D_{B}^{-1} \hat{\bm{b}} - \beta \mu D_{B}^{-1} \hat{A} S^{-1} \bm{1} \\ &\quad \ + D_{B}^{-1} \hat{A} D^{2} (\bm{c} - A^{T} \bm{y} - \bm{s}).
\end{align*}

Solving \eqref{equation:ModifiedNormalEquationSystem} inexactly with inexact solution $\widetilde{\bm{z}}$ and residual $\hat{\bm{r}}$, and calculating $\bm{v} = (D_{B} \hat{\bm{r}}, \bm{0})$ as well as the updates $\widetilde{\Delta \bm{y}} = (D_{B}^{-1} A_{B}^{-1})^{T} \widetilde{\bm{z}}$, $\widetilde{\Delta \bm{s}} = \bm{c} - A^{T} \bm{y} - \bm{s} - A^{T} \widetilde{\Delta \bm{y}}$, and the primal update $\widetilde{\Delta \bm{x}} = \beta \mu S^{-1} \bm{1} - \bm{x} - D^{2} \widetilde{\Delta \bm{s}} - \bm{v}$, the obtained updates now satisfy \eqref{equation:NewtonSystem1} and \eqref{equation:NewtonSystem2}, while the residual exclusively appears in \eqref{equation:NewtonSystem3} \cite[Lemma 4.1]{Mohammadisiahroudi2024EfficientUseOfQuantumLinearSystemAlgorithmsInInexactInfeasibleIpmsForLinearOptimization}.

As an alternative to the MNES for calculating feasible Newton steps, Mohammadisiahroudi \textit{et al.}~\cite{Mohammadisiahroudi2025AnInexactFeasibleInteriorPointMethodForLinearOptimizationWithHighAdaptabilityToQuantumComputers} proposed to precompute bases for the null and row space of $A$ to use respective linear systems for basis coefficients in place of the primal and dual feasibility constraints.
While a basis of $A$'s row space is, by definition, given by its rows or, equivalently, by the columns of $A^{T}$, determining a basis of $A$'s null space (or kernel) requires some additional preprocessing.
Namely, let $B$ and $A_{B}$ as before, and let $A_{N} \coloneqq [a_{j}]_{j \in [n] \setminus B}$.
Now define the matrix $V \in \R^{n \times (n - m)}$ via
\begin{align}\label{equation:NullSpaceMatrix}
    V \coloneqq
    \begin{bmatrix}
        A_{B}^{-1} A_{N} \\
        -\one_{n - m}
    \end{bmatrix},
\end{align}
where $\one_{n - m}$ is the unit matrix of dimension $n - m$.
Then, by construction, the columns of $V$ form a basis of $A$'s null space \cite[Lemma 2.1]{Mohammadisiahroudi2025AnInexactFeasibleInteriorPointMethodForLinearOptimizationWithHighAdaptabilityToQuantumComputers}.
Therefore, $A \Delta \bm{x} = 0$, i.e., $\Delta \bm{x}$ lies in $A$'s null space, if and only if there exists a coefficient vector $\bm{\lambda} \in \R^{n - m}$ so that $\Delta \bm{x} = V \bm{\lambda}$.
Using this representation for $\Delta \bm{x}$ and $\Delta \bm{s} = - A^{T} \Delta \bm{y}$ in \eqref{equation:NewtonSystem3} yields the orthogonal subspace system (OSS)
\begin{align}\label{equation:OrthogonalSubspaceSystem}
    O \bm{w} = \bm{\tau},
\end{align}
where
\begin{align*}
    O &\coloneqq [-X A^{T}\ \ S V] \in \R^{n \times n} \text{ and} \\
    \bm{\tau} &\coloneqq \beta \mu \bm{1} - X \bm{s} \in \R^{n}.
\end{align*}
The first $m$ components of an (inexact) solution $\widetilde{\bm{w}}$ to \eqref{equation:OrthogonalSubspaceSystem} are used to calculate the update $\widetilde{\Delta \bm{s}}$ via multiplication with $- A^{T}$.
Hence, even an inexact update is guaranteed to lie within the row space of $A$.
Analogously, the last $n - m$ components of $\widetilde{\bm{w}}$ constitute the null space coefficients of $\widetilde{\Delta \bm{x}}$.
The latter is obtained by multiplication with $V$, thus also guaranteed to be contained in $A$'s null space.

The skeleton of a primal-dual path-following IPM with any of the aforementioned methods to obtain the Newton steps is summarised in \hyperref[algorithm:InteriorPointMethod]{Algorithm 1}.
The parameter $\mu$ determines to which extent the IPM should iteratively close the duality gap, i.e., how accurately it should satisfy the complementary slackness condition.
Note that this skeleton hides several important hyperparameter which have to be fine-tuned in order to guarantee convergence of the respective \texttt{IPM}-variant, such as the distance of the initial point to the CP, upper bounds on possible feasibility violation, and an enforcing parameter to bound the norm of the residuals arising in inexactly solving the intermediate linear systems.
Especially for infeasible variants, the step length $\alpha$ requires thorough selection to not amplify infeasibilities.
In the benchmark below, I instantiate the Newton system at a canonical strictly positive infeasible-start iterate.

\begin{algorithm}[!ht]
    \caption{\label{algorithm:InteriorPointMethod}\texttt{IPM}($A$, $\bm{b}$, $\bm{c}$, $\bm{x}_{0}$, $\bm{y}_{0}$, $\bm{s}_{0}$, $\mu$)}
    $(\bm{x}, \bm{y}, \bm{s}) \leftarrow (\bm{x}_{0}, \bm{y}_{0}, \bm{s}_{0})$\;
    \While{$\bm{x}^{T} \bm{s} > n \mu$}{
        $(\widetilde{\Delta \bm{x}}, \widetilde{\Delta \bm{y}}, \widetilde{\Delta \bm{s}}) \leftarrow$ solve \eqref{equation:ModifiedNormalEquationSystem} or \eqref{equation:OrthogonalSubspaceSystem} (inexactly)\;
        choose step length $\alpha \in (0, 1)$\;
        $(\bm{x}, \bm{y}, \bm{s}) \leftarrow (\bm{x}, \bm{y}, \bm{s}) + \alpha (\widetilde{\Delta \bm{x}}, \widetilde{\Delta \bm{y}}, \widetilde{\Delta \bm{s}})$\;
    }
    \Return{$(\bm{x}, \bm{y}, \bm{s})$}
\end{algorithm}

\subsection{\label{subsection:QuantumLinearSolversAndInteriorPointMethods}Quantum linear solvers and interior point methods}

Conceptually, all existing proposals of quantum interior point methods (QIPMs) outsource solving the linear system in \hyperref[algorithm:InteriorPointMethod]{Algorithm 1} (line \texttt{3}) to a quantum linear solver algorithm (QLSA) and subsequent tomography, while still conducting the updates and system formulation classically.
In the following, I will introduce basic quantum computing concepts, which are essential for understanding how a linear system is formulated on a quantum device in the first place, and which quantum methods are suitable for solving it.
However, I will not cover all the technical details of the various QLSAs, but refer instead to their respective original proposals and an excellent recent survey of quantum linear system solvers by Morales \textit{et al.}~\cite{Morales2024QuantumLinearSystemSolversASurveyOfAlgorithmsAndApplications}.

A $d$-dimensional quantum system is represented by the complex Hilbert space $\hil = \C^{d}$ and its unit vectors---usually written as \emph{kets} $\ket{\psi} \in \hil$---correspond (non-uniquely) to valid (pure) quantum states.
The smallest non-trivial quantum system is given by the \emph{qubit} with associated Hilbert space $\C^{2}$.
The Hilbert space associated to an $n$-qubit system is given by the $n$-fold tensor product of $\C^{2}$, thus isomorphic to $\C^{2^{n}}$.
Therefore, the Hilbert space's dimension grows exponentially with the number of qubits, analogously to how the number of bit strings grows exponentially with their length, i.e., with the number of bits.
However, while the number of bits required for storing a $d$-dimensional real vector $\bm{v} \neq \bm{0}$ (in floating point representation) is proportional to $d$, the normalised vector $\bm{v} / \norm{\bm{v}}$ can be realised as a quantum state in a $\lceil \log_{2}(d)\rceil$-qubit system.
That is, the entries $v_{i}$, $i \in [d]$, are encoded as amplitudes in some fixed Hilbert space basis $\{\ket{i}\, \vert\, i \in [d]\}$, referred to as the \emph{computational basis}:
\begin{align}\label{equation:AmplitudeEncoding}
    \ket{\bm{v}} \coloneqq \tfrac{1}{\norm{\bm{v}}} \sum_{i = 1}^{d} v_{i} \ket{i}.
\end{align}

Given a vector $\bm{b} \in \R^{d} \setminus \{\bm{0}\}$ and an invertible matrix $M \in \R^{d \times d}$, I can now pose the quantum linear system problem of (approximately) preparing the state $\ket{\bm{x}}$, where $\bm{x} \coloneqq M^{-1} \bm{b}$, which is precisely the task tackled by QLSAs.
In this work, I focus on the class of functional QLSAs---such as the pioneering HHL algorithm---which aim at directly implementing (an approximation to) $M^{-1}$ and applying it to $\ket{\bm{b}}$.
Among the class of functional QLSAs, a recent study~\cite{Lefterovici2026BeyondAsymptoticScalingComparingFunctionalQuantumLinearSolvers} has identified an approach based on Chebyshev series representation~\cite{Childs2017QuantumAlgorithmForSystemsOfLinearEquationsWithExponentiallyImprovedDependenceOnPrecision} as the best-performing in practice.
The underlying idea of this approach is to approximate the function $[x \mapsto x^{-1}]$ outside of a bounded interval around zero by a linear combination of Chebyshev polynomials $\mathcal{T}_{n}$ of the first kind\footnote{The Chebyshev polynomials of the first kind are defined via $\mathcal{T}_{0} \equiv 1$, $\mathcal{T}_{1} = \id$, and $\mathcal{T}_{n + 1} = 2 \id \mathcal{T}_{n} - \mathcal{T}_{n - 1}$ for all $n \in \N$.
Equivalently, they are the unique polynomials satisfying $\mathcal{T}_{n}(\cos(\theta)) = \cos(n \theta)$ for all $n \in \N_{0}$.} and to extend this decomposition via functional calculus to (approximately) implement $M^{-1}$.
Specialised quantum walks use queries to $M$'s non-zero elements to probabilistically implement the (generally non-unitary) operator-valued Chebyshev polynomials $\mathcal{T}_{n}(M)$, $n \in \N$.
Their weighted sum, i.e., the approximation of $M^{-1}$, can then be implemented via the framework of linear combinations of unitaries~\cite{Childs2012HamiltonianSimulationUsingLinearCombinationsOfUnitaryOperations}.

\begin{figure*}[!t]
\centering
\begin{tikzpicture}[
    box/.style={
        rectangle,
        draw=black,
        thick,
        minimum height=0.9cm,
        text width=2.2cm,
        align=center,
        font=\small,
    },
    assumption/.style={
        font=\scriptsize,
        text=black!70,
        text width=2.2cm,
        align=center,
    },
    arr/.style={->, thick, >=stealth},
    barr/.style={->, thick, >=stealth, dashed, black!50},
]

% Main pipeline boxes -- left to right
\node[box] (lp) {LP instance\\[-1pt] {\scriptsize $(A, \bm{b}, \bm{c})$}};

\node[box, right=0.69cm of lp] (newton) {Newton system\\[-1pt] {\scriptsize MNES \eqref{equation:ModifiedNormalEquationSystem} or OSS \eqref{equation:OrthogonalSubspaceSystem}}};

\node[box, right=0.6cm of newton] (params) {system parameters\\[-1pt] {\scriptsize $d$, $s$, $\kappa$}};

\node[box, right=0.6cm of params] (qlsa) {QLSA query count\\[-1pt] {\scriptsize $\mathcal{Q}(s, \kappa, \varepsilon)$}};

\node[box, right=0.6cm of qlsa] (tomo) {tomography cost\\[-1pt] {\scriptsize $\tfrac{d-1}{\varepsilon^2} \cdot \mathcal{Q}$}};

\node[box, right=0.6cm of tomo] (cycles) {quantum cycle\\[-1pt] lower bound};

% Arrows
\draw[arr] (lp) -- (newton);
\draw[arr] (newton) -- (params);
\draw[arr] (params) -- (qlsa);
\draw[arr] (qlsa) -- (tomo);
\draw[arr] (tomo) -- (cycles);

% Benevolent assumptions -- below
\node[assumption, below=0.55cm of qlsa] (a1) {
    1 oracle = 1 cycle\\
    $n_{\text{QAA}} = 1$};
\draw[barr] (a1) -- (qlsa);

\node[assumption, anchor=south] (a2) at (tomo.south |- a1.south) {
    $\varepsilon = 10^{-1}$\\(IR in 1 step)};
\draw[barr] (a2) -- (tomo);

\node[assumption, anchor=south] (a3) at (cycles.south |- a1.south) {
    IPM converges\\in 1 iteration};
\draw[barr] (a3) -- (cycles);

\coordinate (bracestart) at ($(a1.south west)-(0.1,0.1)$);
\coordinate (braceendx)  at ($(a3.south east)+(0.1,-0.3)$);

\draw[decorate, decoration={brace, amplitude=5pt, mirror}, thick, black!50]
    (bracestart) -- (bracestart -| braceendx)
    node[midway, below=7pt, font=\scriptsize, text=black!70]
    {benevolent assumptions};

\end{tikzpicture}
\caption{\label{figure:LowerBoundPipeline}
    Pipeline for computing the quantum runtime lower bound.
    Each dashed arrow marks a benevolent assumption that reduces the bound, making it more favourable to the quantum method.
    If this lower bound already exceeds the classical runtime, no practical quantum advantage is possible.
    The dimension $d$ and the sparsity $s$ are calculated exactly, the condition number $\kappa$ is lower-bounded.
    The QLSA is assumed to return the approximate solution with certainty, thus allowing to set the number of QAA iterations to $n_{\text{QAA}} = 1$.
    Furthermore, the number of IR steps for improving the initial precision $\varepsilon$ is benevolently set to $1$.
    Lastly, the entire IPM is assumed to converge after one iteration, which constitutes another highly benevolent assumption.
}
\end{figure*}
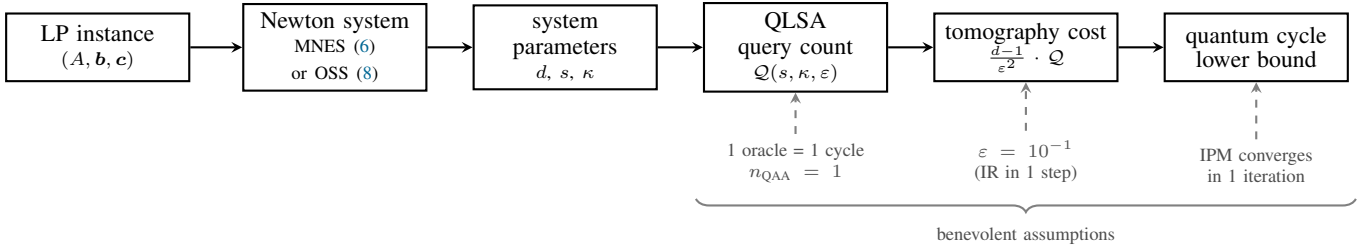

This process is inherently inexact and---by its non-unitary nature---probabilistic.
At a high level, one iteration yields a state $\ket{\eta} \coloneqq \ket{0} \ket{\psi} + \ket{1} \ket{\phi} \in \C^{2} \otimes \C^{d}$, where $\ket{\phi}$ is an approximation to $\ket{M^{-1} \ket{\bm{b}}}$ and $\ket{\psi}$ is some garbage state.
Intuitively, the extra qubit register serves as a flag, indicating whether applying the approximation to $M^{-1}$ was successful ($\ket{1}$) or not ($\ket{0}$).
The success probability of this algorithm is therefore given by $\abs{(\bra{1} \otimes \one_{d}) \ket{\eta}}^{2}$.
A standard technique to boost this success probability is to apply quantum amplitude amplification (QAA)~\cite{Brassard2002QuantumAmplitudeAmplificationAndEstimation} with a simple Pauli-\texttt{Z} operator on the flag qubit as quantum oracle.
Inspired by Ambainis' variable-time QAA~\cite{Ambainis2010VariableTimeAmplitudeAmplificationAndAFasterQuantumAlgorithmForSolvingSystemsOfLinearEquations}, Childs \textit{et al.} incorporated a similar technique into their algorithm.
In essence, the refined version approximates the function $[x \mapsto x^{-1}]$ separately on different sections $[-1, -\lambda] \cup [\lambda, 1]$, $\lambda \in (0, 1)$, where the approximation is cheaper to implement the higher the $\lambda$-value is.
The success probabilities of implementing the approximations, in turn, is larger for smaller $\lambda$-values so that these branches incur lower QAA overheads.

The output of any QLSA (after post-processing on successful runs) is a $d$-dimensional quantum state $\ket{\phi}$ so that $\norm{\ket{\phi} - \ket{M^{-1} \bm{b}}} \leq \varepsilon$, not a classical description of this state.
The latter requires state tomography protocols for determining the computational amplitudes of $\ket{\phi}$ up to some target precision, say also $\varepsilon$.
In the runtime analysis below, I assume that the tomography stage receives only repeated successful preparations of the heralded QLSA output state and is not granted efficient access to a fixed state-preparation unitary for that final normalised state, nor to its inverse or controlled versions; accordingly, I analyse tomography in the copy-access model.
Without further structure, such copy-based tomography protocols scale necessarily with the system size $d$ and with $\varepsilon^{-2}$, and return the computational amplitudes $\expval{i | \phi}$ up to global phase and additive error $\varepsilon$, again subject to a certain success probability.
When tomography is given a fixed state-preparation unitary and its inverse, the query complexity can be improved, at least asymptotically, to scale as $d / \varepsilon$ instead~\cite{VanApeldoorn2023QuantumTomographyUsingStatePreparationUnitaries}.
In any case, the QLSA and the tomography step depend unfavourably on the target precision $\varepsilon$.
As a recourse, Mohammadisiahroudi \textit{et al.}~\cite{Mohammadisiahroudi2024EfficientUseOfQuantumLinearSystemAlgorithmsInInexactInfeasibleIpmsForLinearOptimization} proposed to wrap the hybrid linear solver into iterative refinement (IR)~\cite{Moler1967IterativeRefinementInFloatingPoint}:
In IR, the initial problem $M \bm{x} = \bm{b}$ is first solved inexactly with low precision, e.g., $\varepsilon = 10^{-1}$, yielding an inexact solution $\widetilde{\bm{x}}$.
Then, the residuals $\bm{r} = \bm{b} - M \widetilde{\bm{x}}$ can be computed with high precision, and are used to formulate another linear system for the correction $\bm{c}$: $M \bm{c} = \bm{r}$.
Solving this system---again inexactly---yields an approximate correction $\widetilde{\bm{c}}$.
This procedure can now be repeated for the adjusted solution $\widetilde{\bm{x}}' \coloneqq \widetilde{\bm{x}} + \widetilde{\bm{c}}$, and so on.

\section{\label{section:Methods}Methods}

A faithful benchmark of QIPMs for relevant instance sizes requires large-scale, error-corrected quantum devices for reliably executing the QLSA subroutines.
Without access to such devices, one can only estimate the actual runtime, either by classical simulation, or---as in this work---via concrete runtime formulas.
These formulas only hold under very benevolent assumptions about the quantum hardware and thus provide lower bounds for the runtime on real, imperfect quantum hardware.
Combining these formulas with benevolent assumptions about the algorithm's performance, e.g., convergence after one step, then yields a rigorous lower bound for the actual runtime.
I will not make any attempt to determine how tight these lower bounds are.
Instead, I will use the lower bounds (of possibly low quality) to rule out practical quantum advantage through existing hybrid QIPMs on a large set of benchmark instances, using the following simple argument:
If the quantum runtime lower bound on a given instance is already worse than the runtime of any classical algorithm solving the same instance on any commodity hardware, then running the actual quantum algorithm will never provide a practical advantage for this instance.
The procedure for obtaining the quantum runtime lower bound is summarised in \hyperref[figure:LowerBoundPipeline]{Fig.~1}.

\subsection{\label{subsection:QuantumHardwareAssumptionsAndQueryFormulas}Quantum hardware assumptions and query formulas}

The quantum core of existing QIPMs are the respective QLSAs.
Lefterovici \textit{et al.}~\cite{Lefterovici2026BeyondAsymptoticScalingComparingFunctionalQuantumLinearSolvers} provided quantitative resource counts for four functional QLSAs in terms of specific quantum oracles, and identified the Chebyshev-based version as the best performing one in their own numerical analysis.
I will state their result, highlight the underlying assumptions, and then introduce additional benevolent assumptions to simplify the resource count even further.

\begin{lemma}\label{lemma:QueryCount}({\cite[Lemma 5]{Lefterovici2026BeyondAsymptoticScalingComparingFunctionalQuantumLinearSolvers}})
    Let $M \in \C^{d \times d}$ be an $s$-sparse, non-singular Hermitian matrix with condition number $\kappa$\footnote{If $M \in \C^{d \times d}$ is not Hermitian, such as the OSS matrix $O$, consider $\widetilde{M} = \begin{bmatrix} 0 & M \\ M^{H} & 0\end{bmatrix} \in \C^{2 d \times 2 d}$ which has the same sparsity and condition number as $M$.}, and let $\bm{b} \in \C^{d}$.
    Preparing a state $\ket{\psi}$ with $\abs{\ket{\psi} - \ket{M^{-1} \bm{b}}} \leq \varepsilon$, using the Chebyshev-based QLSA requires at least
    \begin{align*}
        \mathcal{Q} = 8 \left\lceil\sqrt{\lceil s^{2} \kappa^{2} \log_{2}\left(\tfrac{s \kappa}{\varepsilon}\right)\rceil \log_{2}\left(\tfrac{4}{\varepsilon} \lceil s^{2} \kappa^{2} \log_{2}\left(\tfrac{s \kappa}{\varepsilon}\right)\rceil\right)}\right\rceil n_{\text{QAA}}
    \end{align*}
    query calls to the oracles
    \begin{align*}
        \mathcal{O}_{F} : \ket{j, l} \mapsto \ket{j, \iota(j, l)} \text{and}\ \mathcal{O}_{M} : \ket{j, k, z} \mapsto \ket{j, k, z \oplus M_{j k}},
    \end{align*}
    where $\iota(j, l)$ is the column index of the $l$-th non-zero element in the $j$-th row, with the quantum amplitude amplification overhead $n_{\text{QAA}}$.
\end{lemma}

The query count established by \autoref{lemma:QueryCount} already assumes noise-free components, i.e., ignores quantum error correction entirely.
Additionally, I will ignore all other algorithmic components, such as the state preparation of $\ket{\bm{b}}$, and impose the highly idealising assumption that the oracles $\mathcal{O}_{F}$ and $\mathcal{O}_{M}$ can be executed in a single quantum processor cycle.
Furthermore, I simply assume the state $\ket{\psi}$ to be implemented with certainty, which sets $n_{\text{QAA}} = 1$, yielding a lower bound for the execution time even when variable-time QAA is used.

In order to obtain a classical description of $\ket{\psi}$ and hence an approximation to $M^{-1} \ket{\bm{b}}$, the QLSA routine (i.e., the state preparation of $\ket{\psi}$) has to be called many times as part of the state tomography.
Even for sparse $M$ and $\bm{b}$, the solution $M^{-1} \bm{b}$, let alone an (arbitrary) approximation of it, need not be sparse.
However, if both $M$ and $\bm{b}$ only have real entries, so will have the solution $M^{-1} \bm{b}$, and imaginary parts of its approximation may, for simplicity, be ignored.
Under the copy-access tomography model assumed here, any tomography protocol, whether textbook Hadamard tests or more sophisticated methods such as classical shadows~\cite{Huang2020PredictingManyPropertiesOfAQuantumSystemFromVeryFewMeasurements}, requires at least $(d - 1) / \varepsilon^{2}$ uses of the state preparation procedure to resolve all $d - 1$ independent real amplitudes to additive error $\varepsilon$~\cite{Haah2017SampleOptimalTomographyOfQuantumStates}, the total number of quantum cycles for solving the LSE $M \bm{x} = \bm{b}$ to error $\varepsilon$ is lower-bounded by
\begin{align}\label{equation:QuantumCost}
    \tfrac{8 (d - 1)}{\varepsilon^{2}} \left\lceil\sqrt{\lceil \gamma^{2} \log_{2}\left(\tfrac{\gamma}{\varepsilon}\right)\rceil \log_{2}\left(\tfrac{4}{\varepsilon} \lceil \gamma^{2} \log_{2}\left(\tfrac{\gamma}{\varepsilon}\right)\rceil\right)}\right\rceil,
\end{align}
where I introduced the (\emph{effective}) \emph{difficulty} $\gamma \coloneqq s \kappa$.

Practical values for $\varepsilon$ would lie below $10^{-6}$ and drastically increase the requires quantum cycles.
However, using the technique of IR, the initial LSE can be solved to much lower precision, e.g., $10^{-1}$ at the expense of solving additional LSEs for improving the precision.
To massively simplify the analysis, I will assume that the IR variant with low precision $10^{-1}$ yields a high-precision solution already after one step.
This means, one can simply insert $\varepsilon = 10^{-1}$ into \eqref{equation:QuantumCost} to obtain a (probably very loose) lower bound to the quantum cycle count.

\subsection{\label{subsection:IdealisedAlgorithmicBehaviour}Idealised algorithmic behaviour}

IPMs, whether quantum or not, are iterative methods.
In each step, they formulate an LSE which depends on the intermediate solution of the previous step.
Classically simulating a QIPM would therefore amount to executing its classical analogue, i.e., solving the emerging LSEs classically and constructing the subsequent LSEs with the obtained data.
While this would potentially give tighter estimates of the actual quantum runtime, it introduces the risk of selecting poor values for the algorithm's hyperparameters, such as the step length $\alpha$, possibly underestimating the algorithm's performance.
Instead, I will simply assume that the QIPM converges after one step, so that the step size and other hyperparameters do not have to be considered at all.
This is again a massive simplification, but it guarantees that the algorithm's performance is never underestimated, which is consistent with the lower bound for the quantum cycle count per iteration.

Besides ensuring a consistent lower bound on the (quantum) runtime of the entire QIPM, the above assumption massively simplifies the benchmark procedure:
Given an LP with constraint matrix $A$ and constraint vector $\bm{b}$, as well as an initial primal-dual tuple $(\bm{x}_{0}, \bm{y}_{0}, \bm{s}_{0})$, construct the MNES \eqref{equation:ModifiedNormalEquationSystem} or OSS \eqref{equation:OrthogonalSubspaceSystem}, record the respective system's sparsity $s$ and condition number $\kappa$, and plug those values, as well as $\varepsilon = 10^{-1}$, into \eqref{equation:QuantumCost} to obtain a provable lower bound on the quantum cycle count.
\hyperref[figure:LowerBoundPipeline]{Fig.~1} summarises the assumptions and procedures outlined in this and the previous subsection.

For the purpose of lower bounding the cost of one Newton-system solve, I initialise the system with primal-dual variables $(\bm{x}_{0}, \bm{y}_{0}, \bm{s}_{0}) = (\bm{1}_{n}, \bm{0}_{m}, \bm{1}_{n})$, where $\bm{1}_{d}$ is the $d$-dimensional vector with all entries set to $1$.
This is a strictly positive, although not necessarily feasible tuple.
It ensures that the diagonal matrices $X$ and $S$ used in the MNES and OSS formulations are invertible and avoids singular or otherwise degenerate system constructions.
These initial values need not constitute a feasible starting point of the LP, and for feasible IPM variants they are not intended to model a valid algorithmic initialisation.
This is immaterial for the present exclusion analysis:
I benevolently assume that the QIPM converges after a single iteration, so the benchmark uses only the cost of solving the resulting first Newton system and never propagates the resulting iterate into subsequent iterations.

\subsection{\label{subsection:ClassicalPreprocessingAndCompetitor}Classical preprocessing and competitor}

While the two previous subsections justify the lower bounding technique, this subsection contains all the technical details regarding data aggregation and curation.
First of all, the benchmark instances are stored in MPS format, which allows specifying inequality constraints (in both directions), equality constraints, and two-sided bounds on variables.
Furthermore, they generally contain redundant constraints which would cause the constraint matrix of the equivalent problem in standard formulation to not have full row rank.
In a first step, I use HiGHS's \texttt{presolve}-method in order to eliminate all redundant constraints and variables.
Subsequently, by introducing a slack/surplus variable for each inequality constraint and two slack variables for each variable with two-sided bound, as well as splitting free variables $x \in \R$ into two non-negative ones $x = x^{+} - x^{-}$ with $x^{+}, x^{-} \geq 0$, the LP is transformed into standard formulation.
Since all redundant constraints have been eliminated, the resulting constraint matrix $A$ is guaranteed to have full row rank.

It is these adjusted LPs in standard form which serve as the actual benchmark instances.
The classical competitor---HiGHS' \texttt{solve}-method---is directly called on these instances, and the runtime is recorded on a single Intel Core i7 11700K with 8\,GB of RAM on Ubuntu 22.04.4 LTS.
Commercial solvers, such as Gurobi or CPLEX, would probably perform even better on most benchmark instances.
Hence, choosing the potentially worse-performing open-source framework HiGHS as the single classical baseline is in favour of the quantum competitors and thus consistent with the lower bound approach for the quantum runtime.
For lower bounding the number of quantum cycles of both QIPM versions, it suffices to lower-bound the difficulty $\gamma$ of the MNES (resp.\ OSS) system matrix.
Both, the MNES matrix $\hat{M}$ and the OSS matrix $O$, require the preselection of a basis of $A \in \R^{m \times n}$, i.e., of a set of $m$ linearly independent columns of $A$.
These columns are found once via sparse QR factorisation and can be used for both QIPM variants.

Even for sparse $A$, the restricted inverse $A_{B}^{-1}$ is generally dense, i.e., of sparsity $m$.
Since $A_{B}^{-1}$ appears directly as a factor within the construction of $\hat{M}$, the latter carries the sparsity $s = m$, even for sparse LP matrix $A$.
The sparsity of $O$ is a bit more nuanced:
The columns of $-X A^{T}$ directly carry the sparsity of the rows of $A$ since $X = \diag(\bm{x})$ is diagonal with fully non-zero diagonal.
$A_{B}^{-1}$ being dense implies that $A_{B}^{-1} A_{N}$---the upper $m \times m$-block of $V$---is also dense.
The lower block $-\one_{n - m}$ contributes one non-zero entry per row and column.
Therefore, the columns of $S V$ generally have $m + 1$ non-zero entries, as $S$ is again diagonal with fully non-zero diagonal.
The first $m$ rows of $O$ are effectively concatenations of columns of $A$ with generally $m$ non-zero entries stemming from the dense $A_{B}^{-1} A_{N}$ block.
The remaining $n - m$ rows are concatenations of $A$'s columns with rows of $\one_{n - m}$, which only adds one additional non-zero element.
These last row terms are always dominated by the $m + 1$ non-zero elements of the columns of $S V$.
In summary, the sparsity $s$ of $O$ is given by the maximum over all these column and row sparsities.

\begin{figure*}[!ht]
\begin{minipage}[t]{0.49\textwidth}
\centering
    \includegraphics[width=\textwidth]{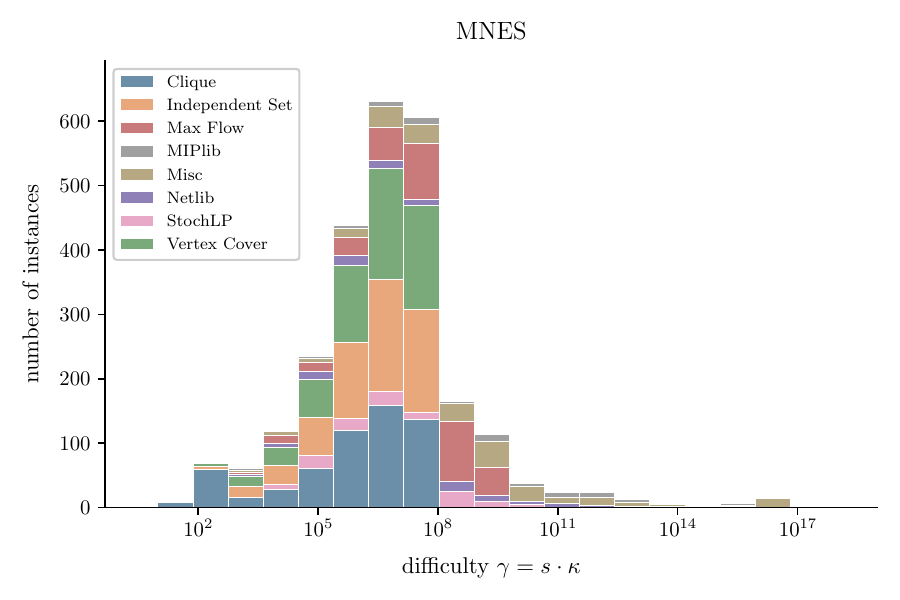}
\end{minipage}
    \hfill
\begin{minipage}[t]{0.49\textwidth}
\centering
    \includegraphics[width=\textwidth]{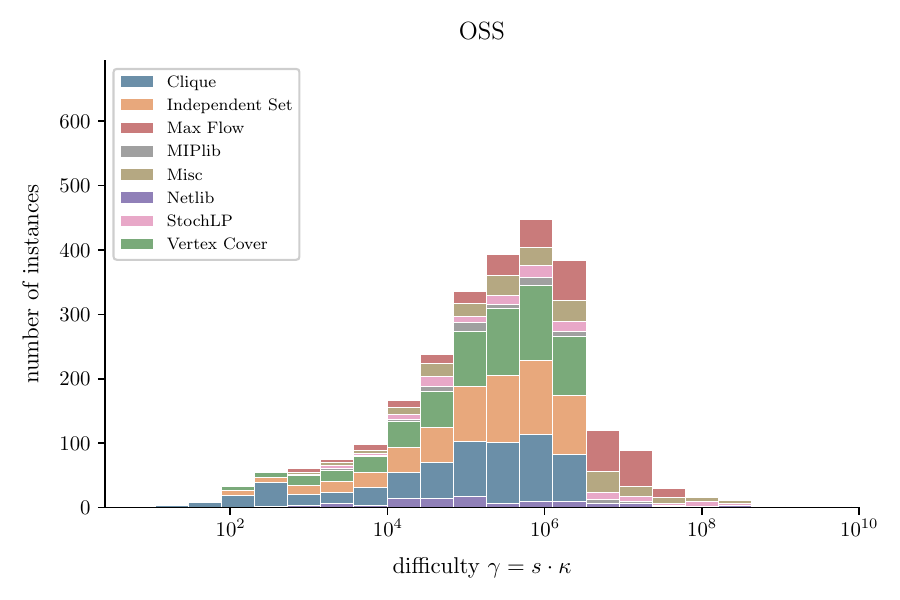}
\end{minipage}
\caption{\label{figure:DifficultyPlots}
    Effective difficulty $\gamma = s \kappa$ of the Newton linear systems across all benchmark instances, for the MNES (left) and OSS (right) formulations.
    The three binary optimisation relaxations---Clique, Independent Set, and Vertex Cover---constitute predominantly easy Newton systems in both cases.
    In fact, the condition number $\kappa$ was often found to be exactly $1$, so that their difficulties are effectively given by their sparsity $s$.
    In contrast, the Max Flow set, consisting of relaxations of an integer optimisation problem, and the benchmark sets MIPib, Misc, Netlib, and StochLP constitute the most difficult Newton systems.
}
\end{figure*}

It remains to describe how the condition numbers $\kappa$ of both system matrices are estimated, yielding lower bounds consistent with the overall goal.
Neither $\hat{M}$ nor $O$ is formed explicitly; instead, both are represented as \texttt{LinearOperator}s---objects that expose only matrix-vector products---which suffices for ARPACK-based iterative eigensolvers and avoids the memory cost of materialising these dense matrices.
For the MNES, the structure $\hat{M} = I + \bar{F}\bar{F}^{\top}$ with $\bar{F} \coloneqq A_{B}^{-1} A_{N}$ is exploited: since $\lambda_{i}(\hat{M}) = 1 + \sigma_{i}(\bar{F})^{2}$, one has
\begin{equation*}
    \kappa(\hat{M}) = \frac{1 + \sigma_{\max}(\bar{F})^{2}}{1 + \sigma_{\min}(\bar{F})^{2}},
\end{equation*}
so it suffices to estimate the extreme singular values of $\bar{F}$ (and, in the special case $n - m < m$ where $\bar{F}\bar{F}^{\top}$ has a null space, $\sigma_{\min} = 0$ and $\lambda_{\min}(\hat{M}) = 1$ exactly).
For the OSS, $\kappa(O) = \sigma_{\max}(O)/\sigma_{\min}(O)$ is estimated directly.
In both cases the same two-stage procedure is applied: $\sigma_{\max}$ is obtained via \texttt{svds("LM")}, whose Ritz values converge from below and therefore underestimate the true value;
$\sigma_{\min}$ is obtained via \texttt{svds("SM")} subject to a $60$-second timeout, whose converged Ritz values are upper bounds by the interlacing theorem.
If the run times out, a fallback evaluates $\min_{w}\ \abs{F \bm{w}}$ over $10000$ random unit vectors $\bm{w}$, which is still an upper bound by the min-max theorem, though potentially a looser one.
Since the numerator is underestimated and the denominator is overestimated, both effects act in the same direction: the estimated condition number is a lower bound on the true one.
The introduced fallback for $\sigma_{\min}$ was necessary, since \texttt{svds("SM")} often did not converge (fast enough) due to the accumulation of many singular values near the minimal one.
Conversely, this clustering effect strengthens the estimation technique via random sampling, since most random unit vectors $\bm{w}$ will result in small values for $\abs{F \bm{w}}$.
However, at this point, a quantitative analysis of this estimate's quality is out of scope;
for the intended exclusion argument, namely for a rigorous lower bound on the quantum cycles, the guarantee of always lower bounding the condition number is sufficient.

\section{\label{section:Results}Results}

I conduct the previously described lower bound analysis on the well-curated benchmark set provided by Ammann \textit{et al.}~\cite{Ammann2023RealisticRuntimeAnalysisForQuantumSimplexComputation}, consisting of the public LP benchmark libraries MIPlib, Misc, Netlib, and StochLP as well as relaxations of Clique, Independent Set, Max Flow, and Vertex Cover on self-generated graph data.
Clique, Independent Set, and Vertex Cover are binary linear optimisation problems, whose continuous LP relaxations have well-conditioned and sparse constraint matrices, hence potentially well-suited for being tackled with QIPMs.
The public benchmark libraries as well as the relaxation of the integer linear problem Max Flow constitute less favourable instances for the QIPM methods.
\hyperref[figure:DifficultyPlots]{Fig.~2} depicts the distribution of effective difficulty among the different instance classes.
It shows that the benchmark set contains both structurally favourable and unfavourable regimes for hybrid QIPMs.
The combinatorial relaxation families mostly yield easy Newton systems, often with condition number close to 1, whereas Max Flow and the public benchmark libraries produce substantially harder systems.

Since the eight LP instance classes are very different in nature, but the instances within a class are structurally similar, I break down the comparison of actual classical runtime and lower-bounded quantum runtime by instance class.
Since the lower bounds on the quantum cycle count are potentially very loose, the only meaningful quantity to observe is the number (or percentage) of instances for which inferred lower bounds on the quantum runtime already lie above the classical runtime.
The lower bound on the quantum runtime is given by multiplying the lower bound on the quantum cycles count with an assumed (logical) quantum cycle duration.
If, for a given instance, the quantum lower bound lies above the classical runtime for realistic (logical) quantum cycle durations, then no practical quantum speed-up can ever be obtained on that instance, as the realistic quantum runtime will always lie above the lower bound.
As a credible lower bound for realistic, logical quantum cycle durations, I chose the current speed record for a physical, entangling two-qubit gate,
$\sqrt{\text{SWAP}}$ on electronic spin qubits in $800\,\mathrm{ps}$~\cite{He2019ATwoQubitGateBetweenPhosphorusDonorElectronsInSilicon}.
This is a fair measure, since the oracles, which are benevolently assumed of constituting a single quantum cycle, are entangling multi-qubit operators.

\begin{figure*}[!ht]
\centering
    \includegraphics[width=1\textwidth]{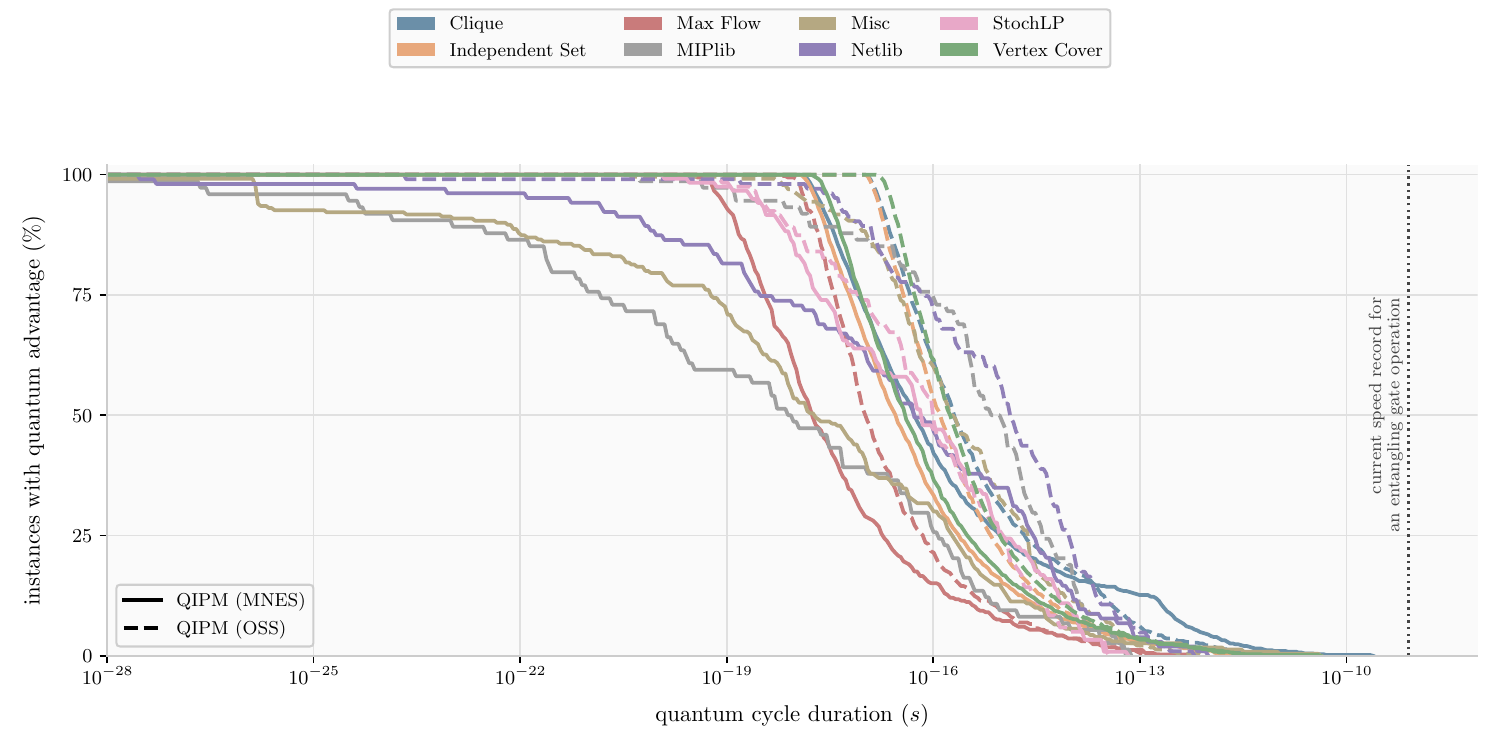}
\caption{\label{figure:ComparisonPlot}
    Percentage of instances (y-axis), divided into eight problem classes, for which the lower bounds of the two QIPM variants' runtimes lie below HiGHS' \texttt{solve} runtime.
    The runtime lower bound is the product of the established lower bound on the cycle count times the assumed quantum cycle duration (x-axis).
    Despite the bounds' looseness due to a chain of several benevolent assumptions, they result in slower quantum runtimes---for both QIPM variants and on all instances---in range of any realistic value for the quantum cycle duration around the current speed record for a physical entangling two-qubit gate at $800\,\mathrm{ps}$~\cite{He2019ATwoQubitGateBetweenPhosphorusDonorElectronsInSilicon}.
}
\end{figure*}

\hyperref[figure:ComparisonPlot]{Fig.~3} shows the percentage of instances for which the quantum runtime lower bound lies below the classical HiGHS runtime, as a function of the assumed quantum cycle duration.
The instances are broken down into the eight instances classes Clique, Independent Set, Max Flow, MIPlib, Misc, Netlib, StochLP, and Vertex Cover.
Despite the looseness of the bounds due to the several benevolent assumptions described above, they result in slower quantum runtimes across all instances for any realistic (logical) quantum cycle duration around the current speed record of $800\,\mathrm{ps}$ for a physical entangling two-qubit gate.
Most importantly, the exclusion result is robust with respect to the choice of Newton-system formulation.
Although MNES and OSS induce different effective difficulties, neither formulation produces a practically relevant regime in which the quantum lower bound undercuts HiGHS at realistic cycle durations.
Since these curves are already obtained under highly benevolent assumptions, the observed lack of advantage should be interpreted as a strong exclusion result rather than as an artefact of pessimistic modelling.

\section{\label{section:Conclusion}Conclusion}

In this work, I have studied hybrid quantum interior point methods (QIPMs) for linear programming (LP) on a broad collection of LP instances.
Equipping the QIPMs with the best-performing functional QLSA and considering both, the modified normal equation system and the orthogonal subspace system formulations, I derived rigorous lower bounds on the quantum runtime under a series of highly benevolent assumptions, including noise-free hardware, oracle calls contributing a single quantum cycle, convergence in one IPM iteration, one step of iterative refinement sufficing for high precision, and numerical lower bounds on instances' condition numbers.
Despite the looseness these assumptions introduce, the resulting quantum runtime lower bounds exceed the runtime of the classical open-source solver HiGHS across virtually all considered instances for any realistic quantum cycle duration around the current speed record of $800\,\mathrm{ps}$ of a physical entangling two-qubit gate.
This rules out practical quantum advantage using current hybrid QIPMs on these \emph{realistic} LP instances, and extends the consistent negative picture established for quantum simplex methods~\cite{Ammann2023RealisticRuntimeAnalysisForQuantumSimplexComputation} and quantum interior point methods for second-order cone programmes~\cite{Dalzell2023EndToEndResourceAnalysisForQuantumInteriorPointMethodsAndPortfolioOptimization} to the setting of QIPMS for LP instances.
The methodology is deliberately one-sided:
The derived lower bounds are only as tight as necessary to rigorously rule out practical quantum runtime advantage on the considered instances.
They generally do not constitute a reliable estimate for the realised QIPM runtime on an actual quantum device.
However, the rigorous lower bounding technique guarantees that actual quantum runtimes will never beat the classical baseline on the considered instances.

Under the assumed copy-access tomography model, the structural source of the hybrid QIPMs' shortcomings is the tomography overhead inherent to the QLSA paradigm.
Extracting a $d$-dimensional, amplitude-encoded classical solution requires at least $8 (d - 1) / \varepsilon^{2}$ repetitions of the QLSA, regardless of the system matrix' spectral properties.
This cost is irreducible by any improvement to the QLSA itself or to the IPM's convergence: even for Newton systems with small condition number, the tomography overhead alone sufficed to push the quantum runtime lower bound above classical runtime.
Hence, the negative results found in this work most likely generalise beyond the benchmark set.
Incurring massive overheads for completely reading out amplitude-encoded vectors is a fundamental issue and equally applies when utilising, e.g., adiabatic instead of functional QLSAs as solver subroutines.
A partial readout or no readout will not suffice to solve the LP in the narrower sense, which, by definition, demands a classical description of the full solution vector.

\section*{Acknowledgment}

I thank Stuart Flannigan, Thomas Kleinert, Andreea-Iulia Lefterovici, Mohammadhossein Mohammadisiahroudi, Andrew Ochoa, and Matthias Wulff for helpful discussions.

\noindent\textbf{Data and code availability statement.}
The depicted data can be found at \url{https://github.com/MarkAureli/qipm}.

%\IEEEtriggeratref{20}
\bibliographystyle{IEEEtran}
\bibliography{IEEEabrv,bibliography}

\end{document}